\documentclass[aps,prl,floatfix,twocolumn,showpacs]{revtex4}
\usepackage{epsfig}
\usepackage{amsmath}
\usepackage{amssymb}

\usepackage{color}

\newcommand{\be}{\begin{equation}}
\newcommand{\ee}{\end{equation}}

\begin{document}


\hsize\textwidth\columnwidth\hsize\csname@twocolumnfalse\endcsname

\bibliographystyle{plain}

\title{Non-linear spin to charge conversion in mesoscopic structures}

\author{Peter Stano$^1$, Jaroslav Fabian$^2$ and Philippe Jacquod$^{3,4,5}$}
\affiliation{
$^1$Institute of Physics, Slovak Academy of Sciences, 845 11 Bratislava, Slovakia\\
$^2$Institute for Theoretical Physics, University of Regensburg, D-93040 Regensburg, Germany\\
$^3$Physics Department, University of Arizona, Tucson, AZ 85721, USA \\
$^4$College of Optical Sciences, University of Arizona, Tucson, AZ 85721, USA \\
$^5$ D\'epartement de Physique Th\'eorique, Universit\'e de Gen\`eve, CH-1205
Gen\`eve, Switzerland}

\vskip1.5truecm
\begin{abstract}
Motivated by recent experiments [Vera-Marun et al., arXiv:1109.5969], we 
formulate a non-linear theory of spin transport in quantum coherent conductors. 
We show how a mesoscopic 
constriction with energy-dependent transmission can
convert a spin current injected by a spin accumulation into an electric signal,
relying neither on magnetic nor exchange fields. 
When the transmission through the constriction is spin-independent, the spin-charge
coupling is non-linear, with an electric signal that is quadratic in the
accumulation. We estimate that gated mesoscopic constrictions 
have a sensitivity that allows to detect 
accumulations
much smaller than a percent of the Fermi energy. 
\end{abstract}
\maketitle

{\it Introduction.}---Spin current detection and measurement protocols are for the most part
based  on ferromagnetic contacts~\cite{Zutic2004:RMP, Fabian2007:APS,johnson1985:PRL,jedema2002b:N,dash2009:N,ciorga2009:PRB,lou2007:N} or 
Zeeman fields~\cite{folk2003:S,frolov2009:PRL,koop2008:PRL,stano2011:PRL}. 
While efficient and well 
controlled, these schemes are not optimized for miniaturization, because 
exchange and magnetic fields have low spatial resolution and because they cannot
detect the weak spin accumulations achievable in 
two dimensional electron gases such as GaAs heterostructures, the 
platform of choice for sub-micron spintronics.
To unleash the full
potential of spintronics at the nanoscale, it is therefore imperative to find novel, 
all-electric protocols. In
sub-micron structures, however, reciprocity and other
symmetry relations constrain the detection of spin currents in the
linear response regime~\cite{zhai2005:PRL, kiselev2005:PRB,adagideli2006:PRL}. In the absence of time-reversal symmetry breaking field
and focusing on two-terminal
geometries, these constraining rules can only be waived by going beyond the linear
response regime. At the nanoscale, this presents important theoretical challenges
as local electric potentials must be determined self-consistently to ensure
gauge invariance~\cite{christen1996:EPL}.

In this manuscript we construct a mean-field non-linear theory
of spin transport through sub-micron scale structures. We use it to
propose a protocol which converts the spin current injected by 
a spin accumulation into a charge signal via the energy-dependent transmission
of a mesoscopic structure. 
While our scheme is general, we focus our discussion on
quantum point contacts and Coulomb blockaded quantum dots,
whose transmission is easily tunable by 
electric gate potentials. We show that the electric response is quadratic in the 
spin accumulation $\delta \mu$ when the transmission is energy-dependent. 
A linear response arises only if the transmission is spin-dependent, which 
usually requires 
an external magnetic field. 
We foresee that our scheme 
has sufficient sensitivity to detect weak spin accumulations 
such as those that can be
generated magnetoelectrically in GaAs heterostructures~\cite{edelstein1990:SSC}, to which the magnetic spin detection schemes are notoriously difficult to apply.

\begin{figure}
\includegraphics[width=0.3\textwidth]{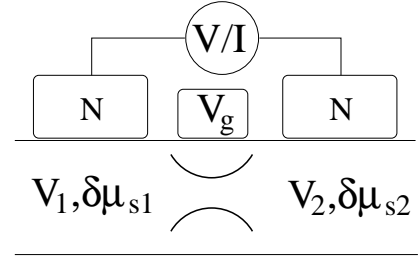}
\caption{\label{fig1} 
 Measurement scheme: A mesoscopic constriction separates two
non-magnetic terminals with spin accumulations $\delta \mu_{s1,2}$. 
The transmission coefficient through the constriction is gate tunable (voltage $V_g$),
and when it is energy-dependent, an electric 
signal that is non-linear in $\delta \mu_{s1,2}$ arises.
\\[-9mm]}
\end{figure}

We consider the standard
measurement setup depicted in Fig.~1. One aims to detect the non-equilibrium spin accumulation drop below the two terminals, $\delta\mu_{s1}\neq\delta\mu_{s2}$, as an electric signal. The spin accumulation origin is not specified, be it ferromagnetic, magnetoelectric, or optical. A recent pioneering work, Ref.~\cite{vera-marun2011:CM}, demonstrated that non-linear effects make the detection possible in graphene even without using ferromagnetic terminals, on which the linear Johnson-Silsbee method relies~\cite{johnson1985:PRL}. A voltage quadratic in the spin accumulation arises due to the energy-dependence of the graphene conductivity near the Dirac point. 
The bottom-line of our theory is that, while
the drift-diffusion approach of  Ref.~\cite{vera-marun2011:CM}  
is appropriate for bulk systems, it
cannot be directly exported to sub-micron structures,
where gauge invariance requires special care~\cite{christen1996:EPL}. 
Furthemore, unlike in graphene, the density of states in GaAs heterostructures is mostly 
energy-independent, thus non-linear effects emerge only if further
constrictions induce energy-dependent transmission $T(E)$. The constriction, such as a Coulomb blockaded quantum dot, a resonant tunneling
barrier, or a quantum point contact (QPC), is the active element in our scheme, converting the spin to charge in proportion to 
$\partial_E T(E)$. This quantity, and this is the crucial point, is fully tunable electrically and independently of the spin accumulation itself, providing our method with versatility necessary for practical spintronics.

{\it Theory calculation.}--We model the detection circuit in Fig.~1 as a quantum scatterer 
connected to two electron reservoirs, each with its own
electrochemical potential and spin accumulation, via two leads.
We start by writing the current in lead $i=1,2$ in the spin subband $\sigma=\uparrow,\downarrow$ (alternatively $\sigma=\pm1$)~\cite{christen1996:EPL}
\be
I_i^\sigma = \frac{e}{h} \int {\rm d}E  \sum_{j\sigma^\prime} \{N_i^\sigma \delta_{\sigma\sigma^\prime} \delta_{ij} - T_{i j}^{\sigma\sigma^\prime} [E, U({\bf r})] \} f_j^{\sigma^\prime}(E).
\label{eq:mother}
\ee
The transmission $T_{ij}^{\sigma\sigma^\prime}$ is the probability that a particle with spin $\sigma^\prime$ injected from reservoir $j$ exits the system
with spin $\sigma$ into reservoir $i$. It depends on the particle energy $E$ and the 
local electrostatic potential $U({\bf r})$.
We consider a spin conserving transmission 
$T^{\sigma\sigma^\prime}\propto \delta_{\sigma\sigma^\prime}$ with
\be
T_{12}^{\sigma\sigma}(E)=T(E) +\sigma \delta T(E) \, .
\label{eq:transmissions}
\ee
For a  spin insensitive structure, the transmission difference is zero, $\delta T=0$, and 
 $T=T^{\uparrow\uparrow}=T^{\downarrow\downarrow}$.
Each lead is characterized by the number $N_i^\sigma$ of transmission channels 
it carries (whose weak energy-dependence we neglect), 
and the particle distribution 
$f^\sigma_i(E) = f(E-\mu_i^\sigma)$,
at the corresponding terminal, 
with the Fermi function
$f(x)=\{\exp[(x-\mu_F)/k_B T]+1\}^{-1}.$
The electrochemical potential of the spin subband $\sigma$, measured from the Fermi energy $\mu_F$, is ($e$ is the electron charge)
\be
\mu_i^\sigma = e V_i + \sigma\, \delta\mu_{si},
\label{eq:potentials}
\ee
where $V_i$ is the applied voltage.

Equation \eqref{eq:mother} is current conserving due to the unitarity condition
$\sum_{j,\sigma} T_{j i}^{\sigma,\sigma'} = N_i^{\sigma'}$.
To guarantee gauge invariance (i.e. that currents are invariant under
an overall voltage shift) one has to take
into account that $U({\bf r})$ is a function of the applied voltages. 
Up to second order in $\mu$'s, the current is~\cite{christen1996:EPL} 
\be \begin{split}
I_1^\sigma &= \frac{e}{h}  \, \int {\rm d}E\,   [-\partial_E f(E)]  \Big\{ 
T_{12}^{\sigma \sigma} (E) ( \mu_1^{\sigma} - \mu_2^{\sigma})\\
& +
(1/2) \partial_E T_{12}^{\sigma \sigma} (E) [( \mu_1^{\sigma})^2 - (\mu_2^{\sigma})^2] \\
& + \int {\rm d}{\bf r}\, [\delta_{U({\bf r})}  T_{12}^{\sigma \sigma} (E)] \delta U({\bf r}) ( \mu_1^{\sigma} - \mu_2^{\sigma} ) \Big\}.
\end{split} 
\label{eq:current 2nd order} \ee
Though the first linear term is explicitly gauge invariant, self-consistent
conditions have to be imposed on $U({\bf r})$ in order to ensure that the non-linear
terms also are gauge-invariant. 
The deviation $\delta U({\bf r})$ of the electrostatic profile from its equilibrium value, 
$ U({\bf r}) = U_{\rm eq}({\bf r}) + \delta U({\bf r})$, results from applied voltages and spin accumulations and the way they are injected into the scattering region.
This deviation can be parametrized by characteristic potentials~\cite{christen1996:EPL}, 
which generally speaking are determined by self-consistent solutions to the 
Schr\"odinger and Poisson equations. To restrain from these (here unnecessary) complications we neglect the spatial dependence of the 
potential changes, $\delta U({\bf r})=\delta U$, and calculate the functional
derivative of $T_{12}^{\sigma \sigma'}$ using the identity
\be
\int {\rm d}{\bf r} \, [\delta_{U({\bf r})} T_{12}^{\sigma \sigma'}(E) ]   = - e  \partial_E T_{12}^{\sigma \sigma'}(E) \, .
\ee
To solve for $\delta U$, we assume a symmetric probe, with equal, spin-independent
coupling to both leads 
\be
e \, \delta U = (\mu_1^\uparrow+\mu_1^\downarrow + \mu_2^\uparrow +\mu_2^\downarrow)/4 \, .
\label{eq:central}
\ee
Using Eqs.~(\ref{eq:transmissions}--\ref{eq:central}) we get our main result, that
the electrical current $I\equiv I_1^\uparrow + I_1^\downarrow$ is
\be
\begin{split}
I=\,&G_1 \, e \, \delta V + G_2 (\delta\mu_{s1}^2-\delta\mu_{s2}^2)  + G_3(\delta\mu_{s1}-\delta\mu_{s2}) \\&+G_4(\delta\mu_{s1}+\delta\mu_{s2}) \, e \, \delta V.
\end{split}
\label{eq:main}
\ee
The formula is explicitly gauge invariant, as it depends only on $\delta V=V_1-V_2$.
The calculation is furthermore current conserving, with $I_2=-I_1$ obtained by substituting
$\delta \mu_{s1} \leftrightarrow \delta \mu_{s2}$ and $\delta V \rightarrow -\delta V$.
We see the emergence of a non-linear spin to charge coupling term $G_2 (\delta \mu_{s1}^2 - \delta \mu_{s2}^2)$, even in 
the absence of any rectification term $\propto \delta V^2$. 
That such a term is absent follows from our
choice of a symmetric potential $\delta U$, in agreement with Ref.~\cite{christen1996:EPL}.
There are four contributions to the current, with conductances
\begin{subequations}
\begin{eqnarray}
G_1 &=& \frac{2e}{h} \int {\rm d}E\,  (-\partial_E f) T(E), \label{eq:G1}\\
G_2 &=& \frac{e}{h} \int {\rm d}E\,  (-\partial_E f) [\partial_E T(E)], \label{eq:G2}\\
G_3 &=& \frac{2e}{h} \int {\rm d}E\,  (-\partial_E f) \delta T(E), \label{eq:G3}\\
G_4 &=& \frac{e}{h} \int {\rm d}E\,  (-\partial_E f) [\partial_E \delta T(E)], \label{eq:G4}
\end{eqnarray}
\end{subequations}
which we discuss in detail below, first for a spin-insensitive, second for a spin-sensitive
constriction.

\begin{figure}
\includegraphics[width=0.5\textwidth]{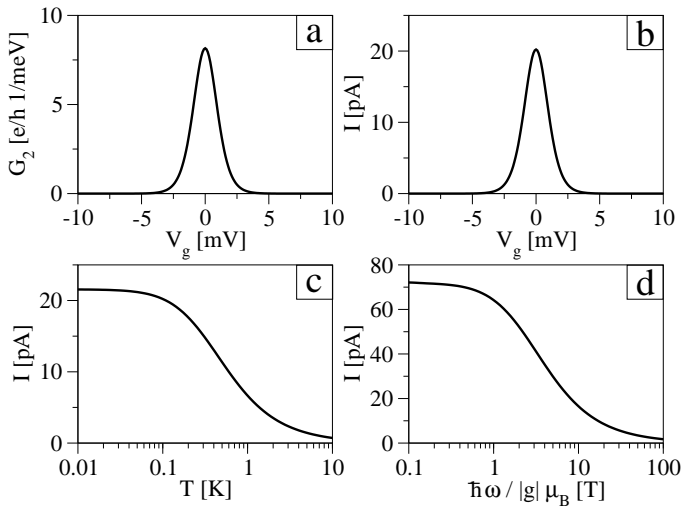}
\caption{\label{fig2} a) Non-linear conductance $G_2$ given in Eq.~\eqref{eq:G2}. b-d) Current, Eq.~\eqref{eq:main} for zero bias, $\delta V$=0, as a function of b) gate voltage determining the QPC transmission [See Eq.~(\ref{eq:QPC 1})],
c) temperature and d) QPC energy resolution in units of the magnetic field.}
\end{figure}

{\it Spin insensitive constriction.}--We first consider $\delta T=0$
in Eq.~(\ref{eq:transmissions}), in which case $G_3=0=G_4$, and focus
our discussion on a gate-defined
QPC in a 2DEG GaAs heterostructure,
with energy-dependent transmission~\cite{buttiker1990:PRB}
\be
T(E)=\left\{ 1+ \exp [-2\pi (E-e\alpha V_g)/\hbar \omega] \right\}^{-1} \, .
\label{eq:QPC 1}
\ee 
The transmission 
is easily tuned by an external gate voltage $V_g$, with a sensitivity set by the QPC
characteristic energy scale
$\hbar\omega$ and $\alpha$ the "lever arm" converting gate voltage into energy. 
We take typical values 
$\hbar \omega=180\,\mu$eV, corresponding to the Zeeman energy of 8 Tesla field at 
the g-factor $g=-0.39$, and $\alpha=0.05$. 
We further fix $\mu_F=8$ meV,  $T=0.1$ K, and spin accumulations $\delta \mu_{s2}=0$, and $\delta\mu_{s1}\equiv\delta\mu_s=0.1$\% $\mu_F$, 
which should be magnetoelectrically achievable~\cite{edelstein1990:SSC}. 

We are now ready to investigate the electric response of the circuit. First, 
we assume that both leads are held at the same potential. Even in this case, 
the spin accumulation
generates a finite current, due to the second term in Eq.~(\ref{eq:main}).
Its magnitude is determined by the non-linear conductance $G_2$ in Eq.~(\ref{eq:G2}), which
we plot in Fig.~2a. It is proportional to $\partial_E T$, and thus maximal when  
the QPC is half open, $T=0.5$. The current is then 
\be
I=G_2 \delta\mu_s^2 \sim \frac{e/h}{{\rm max}(k_B T, \hbar \omega/2\pi)} \delta \mu_s^2,
\label{eq:signal current 1}
\ee
and we plot it in Fig.~2b. 
For our choice of parameters, the current is of the order of tens of pA, which is well above the experimental detection limit. The dependence of the 
current signal on ${\rm max}(k_B T, \hbar \omega/2\pi)$ is demonstrated in Fig.~2c and d. Decreasing $k_B T$ at fixed $\hbar \omega$ ($\hbar \omega$ at fixed $k_B T$), 
the signal first increases before it saturates
when $k_B T \simeq \hbar \omega$.

Alternatively, terminal 2 can be operated as a floating probe. In this case, a finite voltage drop develops, which we find by setting $I_1=0$. One obtains 
\be
e \, \delta V = -G_2 \delta\mu_s^2 / G_1 \, .
\label{eq:signal voltage 1}
\ee
We see that the current is converted into a voltage by the linear conductance $G_1$, given in Eq.~\eqref{eq:G1} and 
plotted in Fig.~3a. 
This agrees with the already mentioned 
absence of a rectification term in our symmetric
QPC~\cite{christen1996:EPL}. We plot the signal voltage in Fig.~3b. In the region where the QPC is closed ($V_g\ll-\hbar\omega$), Eq.~\eqref{eq:signal voltage 1} gives an unphysical 
saturation of $\delta V$ (dashed line). To remove this artifact, we add a small constant to $G_1$, which enforces that $\delta \mu_{s1}$ does not influence $V_2$  
if the QPC is closed. Then the electric signals (current or voltage), in the two 
protocols just discussed behave similarly, 
$I,V \propto G_2 \delta \mu_s^2$. For spin-insensitive constrictions, we see that the electric 
response is quadratic in the spin accumulation.

This result is not specific to a QPC, which we next  replace by a Coulomb-blockaded quantum 
dot. Neglecting inelastic processes
and near resonance, its low-temperature transmission is given by~\cite{stone1985:PRL}
$T(E) = \Gamma^{(1)} \Gamma^{(2)}/[(E-E_0)^2/\hbar^2+(\Gamma/2)^2]$, with the
tunneling rates $\Gamma^{(1)}$ and $\Gamma^{(2)}$ of the resonant level to the left and right leads,
$\Gamma=\Gamma^{(1)}+\Gamma^{(2)}$ and the
resonance peak position $E_0$. At
$E-E_0= \pm \hbar \Gamma/2 \sqrt{3}$, 
$\partial_E T(E)$  takes its maximal value $\pm 9 \Gamma^{(1)} \Gamma^{(2)}/\sqrt{3} \Gamma^3$. Because $T(E)$ 
differs from its QPC expression, Eq.~(\ref{eq:QPC 1}), in its
energy-dependence, the  
shape of the quantities plotted in Fig.~\ref{fig2} will be different; most notably, the signal changes sign upon crossing the resonance.
However, the maximal
current magnitude is still given by 
Eq.~(\ref{eq:signal current 1}) with $\Gamma$ replacing $\omega$.

\begin{figure}
\includegraphics[width=0.5\textwidth]{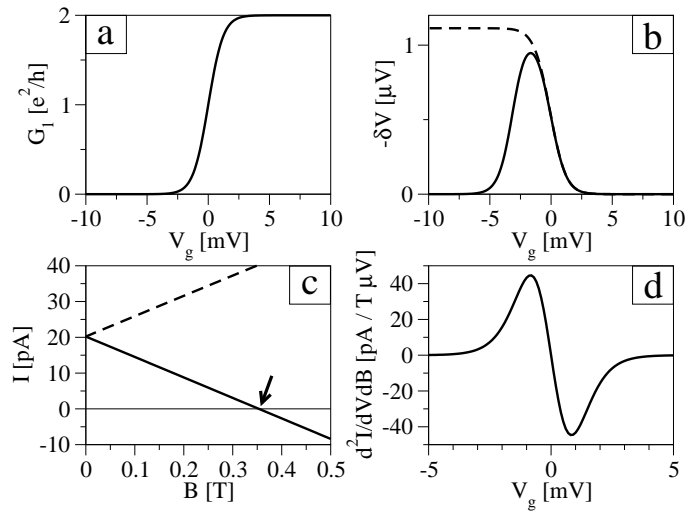}
\caption{\label{fig3} a) Linear conductance $G_1$, Eq.~\eqref{eq:G1}. b) Voltage, calculated according to Eq.~\eqref{eq:signal voltage 1} (dashed line) and adding a constant 0.1 $e^2/h$ to $G_1$ (solid line). c) Current in external magnetic field antiparallel (solid line) and parallel (dashed line) to the spin polarization direction. The arrow denotes the zero current position, the thin line is a guide to the eye. d) Second derivative of the current with respect to $V$ and $B$. }
\end{figure}

{\it Spin sensitive constriction.}--We next consider the case when $\delta T \ne 0$
in Eq.~(\ref{eq:transmissions}), when
the QPC is made spin sensitive, e.g. by an external 
Zeeman field. We assume that the field is parallel to the spin accumulation $\delta \mu_{s1}$
and that it is sufficiently weak that the latter
is not influenced. Equation \eqref{eq:QPC 1} becomes
\be
T^{\sigma\sigma}(E)=\left\{ 1+ \exp [-2\pi (E-\sigma \mu B-e\alpha V_g)/\hbar \omega] \right\}^{-1} \, ,
\label{eq:QPC 2}
\ee 
where the Zeeman energy $\mu B$ 
is added/subtracted from the electron's energy depending on its spin. 
Here, $B$ is the magnetic field, $\mu=(g/2) \mu_B$ and $\mu_B$ is the Bohr magneton.
To linear order in $B$, we then have $\delta T = -\mu B\, \partial_E T$.
We see that a term linear in the spin accumulation has appeared, whose magnitude is 
given by the conductance $G_3$, Eq.~\eqref{eq:G3}. This is a linear response term, similar to the one reported in 
Ref.~\cite{stano2011:PRL}, giving an 
odd magnetoresponse
of the electric current through a QPC in the presence of a spin current.
This term allows to find the sign of the spin accumulation, which is impossible for a spin insensitive probe. This is demonstrated in Fig.~3c, where we plot the current as a function 
of the magnetic field. Applying the field antiparallel (for $g<0$ as in GaAs heterostructures) 
to the spin polarization direction, the Zeeman energy penalty compensates for larger transmission at higher energies. The compensation is exact (the current becomes zero), if 
\be
\delta\mu_s = g \mu_B B_c \, .
\ee
Remarkably, determining the compensation field $B_c$ alone 
allows to measure both the magnitude
and direction of the
spin accumulation. 

For spin-sensitive constrictions, the conductance $G_4$ gives an interesting contribution to the current, which is coupled to the average spin accumulation $\delta \mu_{s1}+\delta \mu_{s2}$
and the voltage bias $\delta V$. We rewrite this contribution as
\be
G_4(\delta\mu_2+\delta\mu_1) \, e \, \delta V = G_4 \Big(\frac{\mu_1^\uparrow+\mu_2^\uparrow}{2} - \frac{\mu_1^\downarrow+\mu_2^\downarrow}{2} \Big) \, e \, \delta V \, ,
\ee
which makes it clear that 
this term describes two different rectification currents in the two spin subbands which are uncompensated if (i) the transport in the subbands happens at different energies, (ii) 
there is a finite bias and (iii) the probe transmission is both spin and energy sensitive. 
Experimentally, this contribution can be identified from the current derivative with respect to both the applied bias and the external field, as in Ref.~\cite{zumbuhl2006:PRL}. We plot this contribution in Fig.~3d where the antisymmetric shape of $G_4\propto \partial_{E}^2 T$, 
strongly contrasts with the symmetric conductances $G_2$, and $G_3$.

{\it Conclusions.}--We have shown how spin accumulations 
can be converted into electric signals in mesoscopic systems with 
energy-dependent, but spin-conserving transmission. When transport
is spin-independent and in the absence of voltage bias, 
the conversion occurs in the non-linear regime,
and the electric signal is quadratic in the spin accumulation. In sub-micron
structures, such nonlinearities have to be treated self-consistently in local
electrostatic potentials generated by the finite applied biases. We did that  within a simplified mean-field approach, which resides in neglecting the spatial structure of the potential changes $\delta U({\bf r})$.  It is reassuring that the non-linear signal arises within this approximation, which restrains from details of the constriction. Further, device specific, spin rectifying effects may arise from the spatial effects in $\delta U({\bf r})$, along the lines of Refs.~\cite{christen1996:EPL} We also note that for no applied bias our approximation becomes exact since $\delta U\to 0$.

In the case of transmission through a QPC, 
Eq.~(\ref{eq:signal current 1}) suggests that it is its
energy resolution, with $\hbar \omega \simeq 2-3 K$
typically, rather than the temperature, which limits the signal magnitude and hence the 
sensitivity of our approach. Alternatively, a Coulomb-blockaded quantum dot 
can be used, where the resolution is given 
by the tunneling width which can easily reach $h \Gamma \simeq 0.1 K$ or less
(see e.g. \cite{patel1998:PRL}).
Close to resonance, we would expect such a quantum dot to enhance
the signal by at least one order of magnitude compared to the results shown in 
Fig.~2d.

{\it Acknowledgements}.--This work was supported by the EU project Q-essence, 
meta-QUTE ITMS NFP 26240120022, and CE SAS QUTE, the NSF under grant
DMR-0706319 and the Swiss Center of Excellence MANEP. J. F. acknowledges
support from the DFG SFB 689 and SPP 1285.

\end{document}